\documentclass[aps,pre,floatfix,tightenlines,longbibliography,notitlepage]{revtex4-1}
\usepackage{graphicx}
\usepackage{bm}
\usepackage[normalem]{ulem}
\usepackage{amsfonts}
\usepackage{color}
\usepackage{verbatim}
\usepackage{amsmath}    
\usepackage{epsfig}
\usepackage[caption=false]{subfig}  
\usepackage[breaklinks,colorlinks = true,linkcolor = blue,urlcolor=blue,citecolor=blue]{hyperref}
\usepackage{amssymb}

\newcommand{\ang}[1]{\left\langle{#1}\right\rangle}

\newcommand{\corr}[1]{\textcolor{black}{{#1}}}

\newcommand{\icts}{International Centre for  Theoretical Sciences, Tata Institute of Fundamental Research,  Bangalore 560089, India}

\begin{document}
\title{Lagrangian Manifestation of Anomalies in Active Turbulence}
\author{Rahul K. Singh}
\email{rksphys@gmail.com}
\affiliation{\icts}
\author{Siddhartha Mukherjee}
\email{siddhartha.m@icts.res.in}
\affiliation{\icts}
\author{Samriddhi Sankar Ray}
\email{samriddhisankarray@gmail.com}
\affiliation{\icts}
\begin{abstract}

	We show that Lagrangian measurements in active turbulence bear imprints
	of \textit{turbulent} and anomalous \textit{streaky} hydrodynamics
	leading to a self-selection of persistent trajectories---\textit{L\'evy
	walks}---over diffusive ones. This emergent dynamical
	heterogeneity results in a super-diffusive \textit{first passage}
	distribution which could lead to biologically advantageous motility.
	We then go beyond single-particle statistics to show that for
	the pair-dispersion problem as well, active flows are at odds with
	inertial turbulence.  Our study, we believe, will readily inform
	experiments in establishing the extent of universality of anomalous
	behaviour across a variety of active flows.

\end{abstract}

\maketitle
\section*{Introduction}
Flowing active matter, such as dense
suspensions of bacteria, has emerged as one of the most intriguing class of
problems in complex, driven-dissipative systems which sits at the intersection
of non-equilibrium statistical physics, biophysics, soft-matter and of course
fluid dynamics~\cite{SR2013,SR2017,doostmohammadi2018active,alert2021active}.
What makes such low Reynolds number systems particularly fascinating is the
emergence of rich and complex collective patterns at scales much larger than, but
driven by, relatively simpler individual
dynamics~\cite{TTS2005,SR2013,JY2014,SR2017} as well as its ubiquity across
systems as diverse as bacterial colonies~\citep{WDY12,DWG13},
suspensions of microtubules and molecular motors~\citep{wu2017transition,
Sanchez12,Sumino12}, or schools of fish~\citep{Katz11} and bird
flocks~\citep{SR2013,SR2010,Cavagna10}. In dense active systems, typically
those involving microscopic entities, the interactions between individual
agents lead to unorganized, often vortical, dynamics with self-similar
distribution of energy across several length scales. This last aspect, namely
the \textit{appearance} of the flow field and the power-laws which emerge in
measurements of the kinetic energy across Fourier modes~\citep{Giomi2012,WDY12,Alert2020,alert2021active}, lead to such states
being called \textit{active turbulence} in analogy with similar traits of high Reynolds number inertial turbulence~\citep{alert2021active}.

However, there are important distinctions between inertial and active
turbulence. The most striking of these being that unlike high Reynolds number
flows, experiments suggest non-universal signatures of L\'evy walks and
anomalous diffusion in measurements of mean-square-displacements (MSDs) in
active
suspensions~\citep{wu2000particle,kurtuldu2011enhancement,morozov2014enhanced,ariel2015swarming}.
These observations were substantiated in a recent theoretical
work~\citep{Sid21} which showed not only the robustness of the anomalous
diffusion and its coincidence with the emergence of novel, \textit{streaky}
structures in the flow for optimal activity, but also why in
earlier theoretical studies~\cite{joy2020friction} such a scaling regime
remained masked. 

The fact that trajectories of tracers in such systems show L\'evy statistics is
consistent with other examples from the natural world where  \textit{active
agents} L\'evy walk or fly~\citep{Klafter87,Dhar2006,Ben94,Krummel16,Cole95}. Nevertheless, 
this throws up interesting questions regarding the nature of trajectories 
and hence, fundamental issues of Lagrangian turbulence in such dense bacterial suspensions. 
In particular, a consequence of this could well be that while for low activities, trajectories are nearly
always meandering (hence diffusive), at higher activities, there is a balance
between those which follow L\'evy statistics and those which do not. In fact,
even for a given tracer it is conceivable that depending on the flow it
experiences, its trajectory could either be persistent or random. 

\corr{We address some of these issues relevant to dense (and dry) bacterial suspensions, by taking a continuum, numerical approach to the problem. The generalized hydrodynamic model (see, also, Refs.~\citep{WDY12,zanger2015analysis} for a detailed description), which is essentially a modified version of the Navier-Stokes equation employing terms that lead to pattern formation (similar to Swift-Hohenberg theory~\citep{swift1977hydrodynamic}) and flocking behaviour (Toner-Tu theory~\citep{TT95,TT98}), is used to simulate the evolution of the coarse-grained velocity field ${\bf u}$ of bacterial suspensions, and is given as:}
\begin{equation}
\partial_t{\bf u} + \lambda {\bf u}\cdot {\nabla\bf u} =-{\bf \nabla}p-\Gamma_0\nabla^2{\bf u}-\Gamma_2\nabla^4{\bf u}-(\alpha + \beta|{\bf u}|^2){\bf u}
\label{GNS}
\end{equation}
\corr{where ${\bf u}$ satisfies the incompressibility constraint $\nabla\cdot{\bf
u}=0$. The parameter $\lambda>1(<1)$ corresponds to pusher (puller)-type
bacteria while the higher order derivatives along with the non-linear self-advective term drive the formation of chaotic flow patterns when $\Gamma_0,\Gamma_2 > 0$. A key difference is that the usual diffusion term of the Navier-Stokes equation appears with the opposite sign, and acts towards inducing a turbulence instability in the bacterial system, while the additional bi-Laplacian term aids dissipation. The last term is a Toner-Tu drive~\citep{TT95,TT98}, which is effectively a quartic Landau-potential for the velocity, with the activity $\alpha$ taking both
positive (friction) and negative (active injection) values, while $\beta$ is strictly positive for stability and causes most of the momentum dissipation. For driven active systems ($\alpha<0$), this term leads to global polar ordering with a velocity $v_0 = \sqrt{|\alpha|/\beta}$.}

\corr{In our simulations, we solve Eq.~\eqref{GNS} by using a de-aliased,
pseudo-spectral algorithm on square, periodic domains of lengths $L\in \lbrace
20,40,80\rbrace$ which are discretized by using $N^2 \in \lbrace 1024^2, 4096^2\rbrace$
collocation points. The simulations are performed for $5\times 10^5$ iterations
with time-steps $\delta t = 0.001$ and, in some cases,  
$\delta t = 0.0002$ for higher temporal resolution (note that a simulation time of $\sim 30$ is $\sim 1$ minute of real time~\citep{WDY12}). The other parameters of the model are taken to be consistent with earlier studies~\citep{MJ1,AJ2020,Sid21,James2021}, and these were carefully chosen to reproduce the physically observed flow patterns and Eulerian statistics~\citep{WDY12}. Here, $\Gamma_0 = 0.045$, which corresponds to $-53$ $\mu m^2/s$, $\Gamma_2 = \Gamma_0^3$ and $\lambda = 3.5$~\citep{WDY12}. The swarming speed of \textit{Bacillus subtilis}, which is linked to the activity, can vary over a wide range of $25-100$ $\mu m/s$, depending on the oxygen concentration of the system and the boundary conditions~\citep{janosi1998onset,WDY12,sokolov2012physical,ariel2015swarming}. These velocities in
the experiment are associated with the typical velocities which arise in the
hydrodynamic description: $v_\Gamma = \sqrt{|\Gamma_0|^3/\Gamma_2}$ or $v_0 =
\sqrt{|\alpha|/\beta}$. An empirical comparison between simulations and experiments (see~\cite{WDY12}, Supporting Information) suggested that $\alpha = -1$ and $\beta = 0.5$ (simulation units) corresponded
to $-0.5/s$ and $4\times 10^{-4} s/\mu m^2$ (physical experimental units),
respectively. These physical units correspond to a velocity scale $\sim 35 \mu
m/s$, with the corresponding simulation velocity scale $v_0 =
\sqrt{|\alpha|/\beta} = \sqrt{2}$. However, as was shown in a recent theoretical study~\citep{Sid21}, phenomena like anomalous diffusion and L\'evy walks, that were recently observed in experiments~\citep{ariel2015swarming}, \textit{only} become robust at sufficiently high activity levels in simulations (around $\alpha = -6$). Therefore, while keeping $\beta=0.5$ fixed, we vary the activity over a wide range $-6 \leq \alpha \leq -1$. We note that the highest value of $\alpha = -6$ yields $v_0 \approx 2.4\sqrt{2}$ (simulation
units) or $v_0 \approx 72\mu m/s$ (choosing, for convenience, the
average velocity in the range $25\mu m/s - 35\mu m/s$ reported by~\cite{WDY12} as reference), which is well within the physically viable range of bacterial velocities~\citep{WDY12,sokolov2012physical}. We caution the reader that this mapping of parameters should be seen as a rough guide, since the calibration of coefficients between theory and experiments is largely empirical.} Lastly, the flow is seeded with $10^5$ randomly distributed
tracers which evolve as $d{\bf x}/dt={\bf u}({\bf x}(t))$, with ${\bf x}(t)$ being the tracer location at time $t$, after a spinup time of $2\times 10^4$ iterations (when the flow reaches a statistically steady state). We use a fourth-order Runge-Kutta scheme, along with a bilinear interpolation 
scheme to obtain the fluid velocity at the particle positions ${\bf u}({\bf x}(t))$, to evolve the tracers \corr{with statistics being stored every $100$ iterations}.

\begin{figure}
\includegraphics[width=1.0\textwidth]{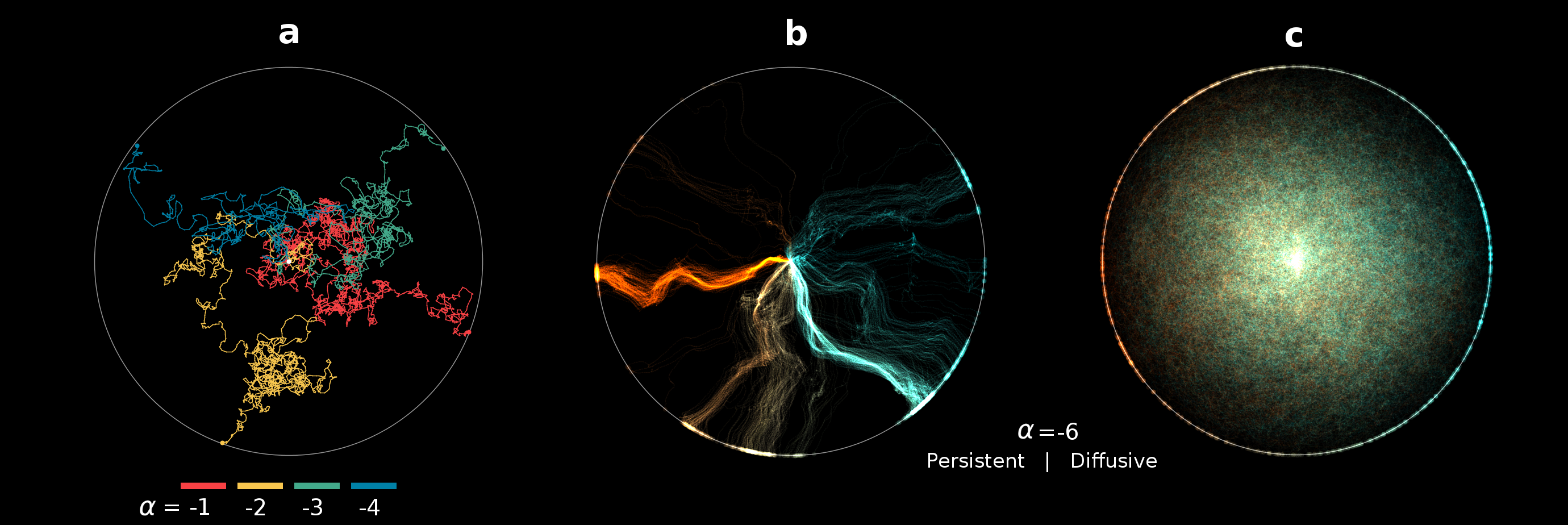}
	\caption{Representative trajectories for (a) suspensions with different degrees of activity \corr{$\alpha$ (see Eq.~\eqref{GNS})} as well as for highly active suspensions 
	showing the (b) fastest 1\% tracers, that are persistent and anisotropic and (c) slowest 1\% tracers that are diffusive and isotropic. 
	Different trajectories are translated 
	to a common origin for clarity and the color variation is a function of the angle at which they reach the bounding circle while the brightness 
	is proportional to the density of trajectories;  see also \url{https://youtu.be/CPJ3ZlXBf-k}. (These images, and Fig.~\ref{RegionalTraj}, were generated with Processing~\citep{reas2007processing,pearson2011generative}).}
\label{Traj}
\end{figure}

In Fig.~\ref{Traj}(a) we show representative trajectories from our simulations with increasing levels of
activity. While for suspensions with low activity ($\alpha = -1$), the particle
motion is predominantly diffusive with large, \textit{knotted} regions of \textit{random-walk}, 
the more active fields ($\alpha = -4$) give rise to trajectories which
have a persistent motion showing characteristic signatures of L\'evy walks; a precise definition 
of the activity parameter $\alpha$ is given below.
However, these L\'evy-like, persistent trajectories are only one part of the
story. Careful measurements in these simulations indicate that even at higher
activity ($\alpha = -6$) it is easy to find in an ensemble 
trajectories that are persistent (Fig.~\ref{Traj}(b)) and those that remain
predominantly diffusive (Fig.~\ref{Traj}(c)); see also \url{https://youtu.be/CPJ3ZlXBf-k}. (We note that the trajectories were artificially moved to a common center for visualization; in 
simulations, their origins are distributed randomly at different points in the flow.)

This leads us to critically examine the nature of trajectories in
such highly active systems and establish connections between anomalies of the
emergent coarse-grained velocity field and its effect on
the resultant Lagrangian statistics. As a result we 
uncover a remarkable dynamical heterogeneity in trajectories, implicit in
Fig.~\ref{Traj}, and its critical role in assisting the swarm for efficient
foraging through first-passage statistics. We also show, in a way which is
easily amenable to experiments, that such persistent motion are facilitated by
the novel, emergent structures in the bacterial
field---\textit{streaks}---which have no known counterparts  in inertial, high
Reynolds number turbulence. We conclude by going beyond single-particle
statistics and investigating the pair-dispersion problem in active turbulence.

Trajectories, especially for high activity, often comprise of 
long walking-segments of varying
step-lengths, interspersed with turning points as seen in Figs.~\ref{Traj}(a) and~\ref{Traj}(b). 
Identifying the ``turns''~\citep{ariel2015swarming,Sid21} is then crucial to segment trajectories into their
constituent step-lengths (and waiting-times) for the analysis that follows. This is done by first calculating a turning angle
$\theta$ at each point along the trajectory at time intervals $\Delta t$ as
$\cos\left(\theta(t)\right) = \frac{ \Delta \mathbf{r}(t)\cdot \Delta \mathbf{r}(t+\Delta t) }{\vert\Delta \mathbf{r}(t)\vert \vert\Delta \mathbf{r}(t+\Delta t)\vert},$
where $\Delta \mathbf{r}(t) = \mathbf{r}(t) - \mathbf{r}(t - \Delta t)$; our results are insensitive for a wide range of $\Delta t$. Walking-segments and turns can be identified using a threshold $\theta(t)>\theta_c$, which in turn gives the step-lengths $d$ and waiting-times $\tau$ of the segments between successive turns. 
We choose $\theta_c = 30^\circ$ for this study~\citep{Sid21} and have checked that our results remain unchanged for $25^\circ \le \theta_c \le 45^\circ$.
 
\section*{Fast and Slow Trajectories and The Role of Emergent Flow Structures}
\begin{figure}
\includegraphics[width=1.0\textwidth]{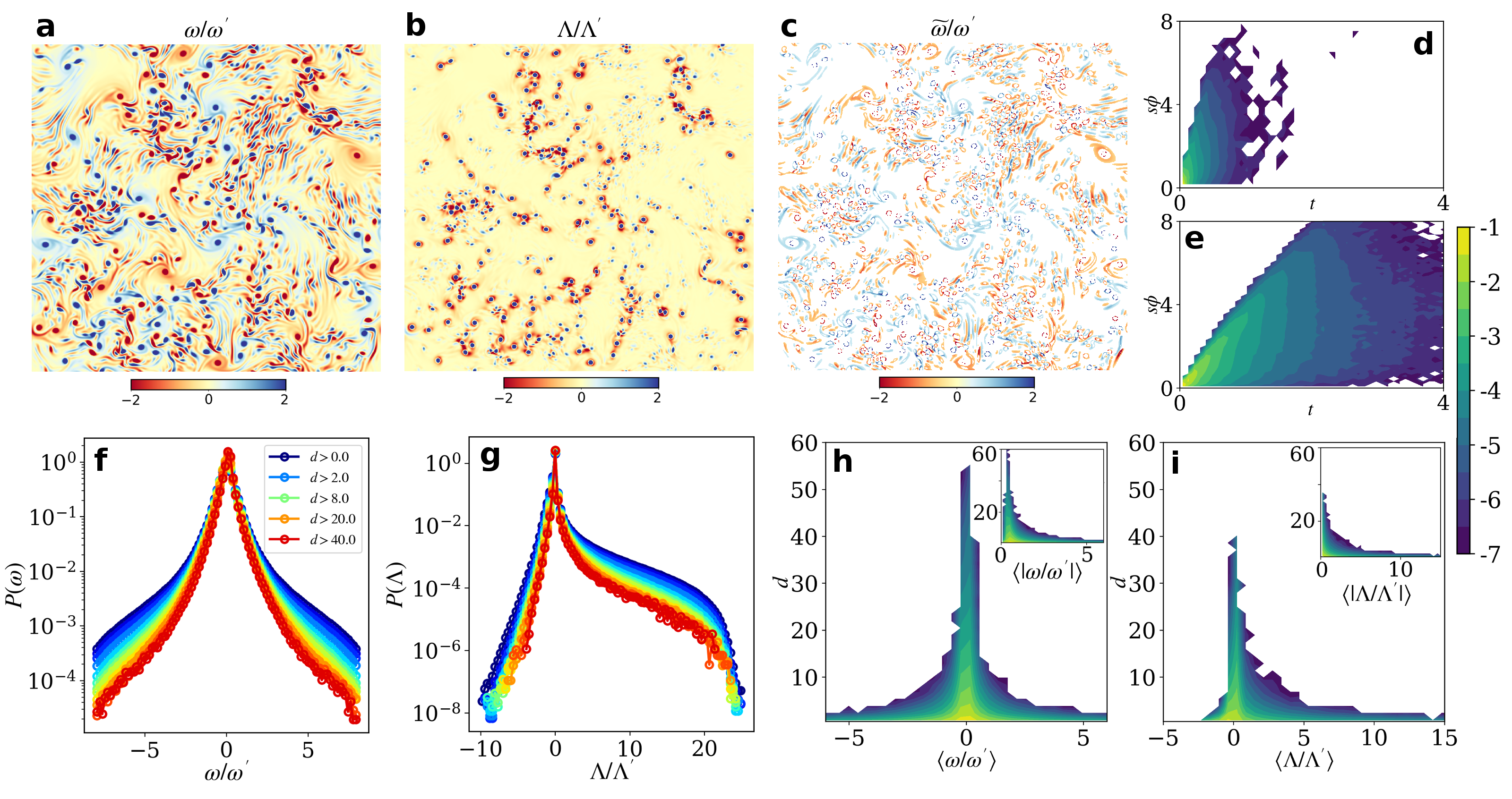}
	\caption{Pseudocolor plots of the (a) vorticity $\omega$, (b) Okubo-Weiss parameter $\Lambda$ and (c) 
	masked vorticity $\widetilde \omega$ fields showing the vortical, straining and streaky patches. Joint-distributions 
	of the residence time and normalised distances of particles (d) within and (e) outside the streaks showing streaks facilitating 
	enhanced displacement in shorter times. 
		Probability distribution functions of the Lagrangian history of (f) vorticity and (g) Okubo-Weiss criterion, conditionally sampled over trajectory segment lengths $d$, show that persistent segments pass preferentially through quiescent regions of $\omega$ and $\Lambda$. 
		Joint distributions of the segment lengths $d$ with (h) $\omega$ and (i) $\Lambda$ (insets show the distribution with 
		$|\omega|$ and $|\Lambda|$) further emphasise that the longer trajectory segments mostly occur in quiescent field regions (the colorbar for the joint distributions shows logarithmically spaced decades).}
\label{ConditionedPdf}
\end{figure}

Do tracers in highly active suspensions ($\alpha = -6$) show a bias in the way they
sample the relatively weaker and stronger vorticity $\omega = \nabla \times {\bf u}$ regions 
of the flow (Fig.~\ref{ConditionedPdf}(a))? The  probability distribution function (pdf) of 
the vorticity field normalised by $\omega' = \sqrt{\left\langle \omega^2
\right\rangle}$, where $\left \langle \cdot \right\rangle$ denotes spatial
averaging, along the Lagrangian trajectories and conditioned on the
step-lengths $d$ shows (Fig.~\ref{ConditionedPdf}(f)) that trajectories are less likely to have
persistent, \textit{directed} motion in regions of strong vorticity: Quiescent
regions favour persistent motion (Fig.~\ref{Traj}(b)).

While this is a useful starting point, it
does not capture the two main geometrical effects---vortical and straining
regions---which characterise (inertial and active) two-dimensional
turbulence. A first approach to this problem is through the 
Okubo-Weiss parameter  $\Lambda = (\omega^2
-\sigma^2)/4\langle \omega^2\rangle$, where $\sigma^2 = S_{ij}S_{ij}$ is the square of the strain-rate $S_{ij} = ( \nabla \mathbf{u} + \nabla \mathbf{u}^T)/2$,
measured along individual trajectories: The sign of the Okubo-Weiss criterion~\citep{okubo1970horizontal,weiss1991dynamics}
determines if the particle is either in a vortical ($\Lambda > 0$) or a straining
($\Lambda < 0$) region as shown in Fig.~\ref{ConditionedPdf}(b) for the Eulerian vorticity field of panel (a).

Figure~\ref{ConditionedPdf}(g)---the Lagrangian pdf 
of $\Lambda$, measured along particle tracks and conditioned on
step-lengths in a manner similar to panel (f)---while clearly showing an
overall preferential bias for vortical regions, brings out the inference drawn
before more clearly. Tracers are far more likely to move in long,
straight segments when they are in regions of the flow where both $\omega$ and
$\Lambda$ have a small magnitude. This bias for persistent motion in quiescent
regions of the flow is further affirmed by the joint distributions of tracer
displacements $d$ with $\omega$ and $\Lambda$ as shown in
Figs.~\ref{ConditionedPdf}(h) and \ref{ConditionedPdf}(i), respectively. The
insets in these panels are the same joint distributions but for 
$|\omega|$ and $|\Lambda|$.  These clearly show
that large displacements are more probable when both $\omega$ and $\Lambda$ are
small: Long straight excursions instead of diffusive meandering occur in these
regions. 

These distributions, however, cannot trace the \textit{origins} of extremely
coherent motion of tracers, illustrated in Fig.~\ref{Traj}(b),
which contribute most to the anomalous diffusive behaviour in highly active
suspensions. This is because a simple decomposition of the flow field in terms
of vortical and straining regions fails to capture a recently
discovered~\cite{Sid21} additional emergent \textit{geometrical} feature (with no counterpart in inertial 
turbulence):
\textit{Streaks} which are striped structures with alternate signs of $\omega$ as seen in 
Fig.~\ref{ConditionedPdf}(a). 

\begin{figure}
\includegraphics[width=1.0\textwidth]{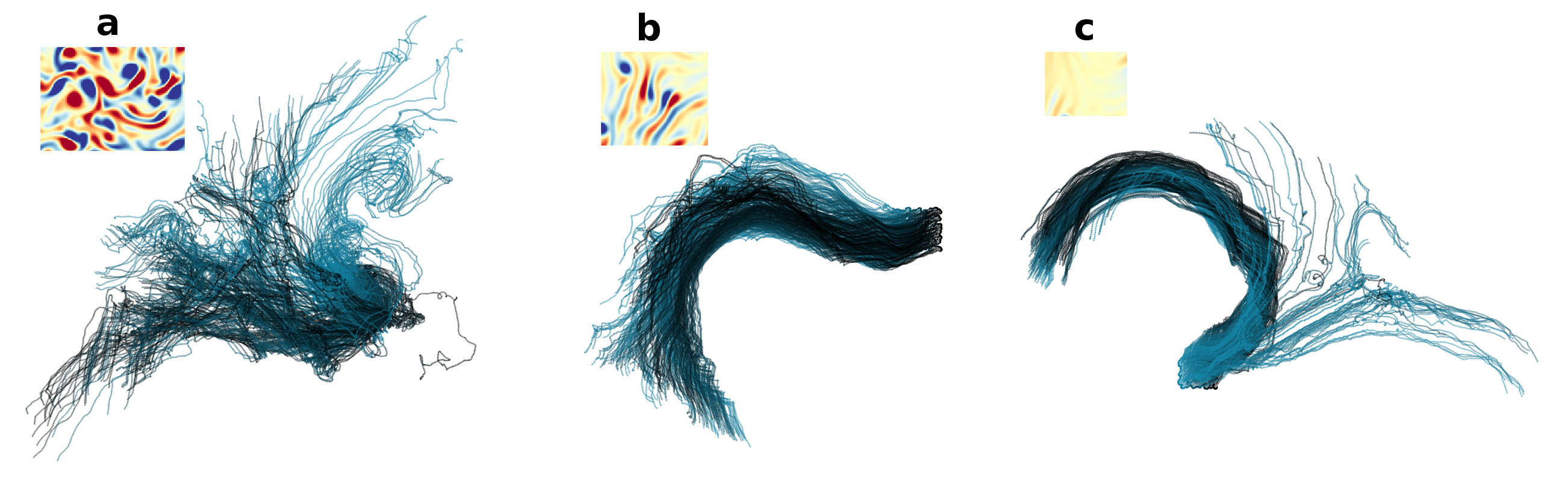}
	\caption{Representative bundles of trajectories originating in 
	(a) vortical spots,  (b) streaks and 
	(c) quiescent regions. While the trajectories spread incoherently 
	in (a) and coherently in (b), there is a mix of both for (c).
	The trajectories are coloured, varying smoothly from blue (left edge)
	to black (right edge), based upon their $x$-coordinates with
	the colour and gives a sense of mixing vis-\`a-vis
	coherence in the trajectory bundles.}
\label{RegionalTraj}
\end{figure}

Are streaks the primary cause for persistent L\'evy walks seen
in our Lagrangian measurements? In fact, the local flow geometry, and how it
evolves in time, essentially governs the fate of a trajectory that passes through it.
We demonstrate this in Fig.~\ref{RegionalTraj}, which shows representative
close-by trajectories originating in (a) vortical, (b) streaky or (c)
quiescent straining regions; the color panels above the trajectories show the
vorticity in the small region of the flow where these trajectories originate.
This immediately brings out the striking difference in the fate of tracers
depending on where they are in the flow. While particles originating in vortical
spots (Fig.~\ref{RegionalTraj}(a)) travel diffusively and incoherently, the ones which start in streaky 
regions (Fig.~\ref{RegionalTraj}(b)) form a coherent bundle with persistent and correlated motions.
Trajectories originating in quiescent regions (Fig.~\ref{RegionalTraj}(c)) show elements of both diffusive
motion with periods of persistence. 

How, then, do we understand this puzzling behaviour? As the underlying field evolves, the trajectories which may have
originated in a particular geometry of the flow are likely to encounter a
different flow topology in time. This rules out 
mean-squared-displacements, conditioned on where trajectories originate, 
as a diagnostic  for a couple of reasons.
Firstly, such an exercise can be only performed up to short times during which
the field remains largely unchanged.  Secondly, these short-time time MSDs are
invariably ballistic, regardless of where they originate.  Hence, the subtle
correlation between the emergent vagaries of the Eulerian field and the
individual trajectories must be found within the Lagrangian history of
particle dynamics.

But first, a sharper definition of what is a streak must be made which goes beyond the 
visual. To that end, careful measurements suggest that streaks coincide 
with places with (a) relatively moderate values of $\omega$ and (b) low
magnitudes of $\Lambda$. This becomes clear upon comparing the vorticity field,
in Fig.~\ref{ConditionedPdf}(a), with the $\Lambda$ field at the same time instance,
shown in Fig.~\ref{ConditionedPdf}(b).
This hints at a criterion, albeit somewhat \textit{ad-hoc},
for identifying the streak regions. We define streaks as 
regions of the flow where, locally, the vorticity is bounded from below and the
Okubo-Weiss criterion is bounded from above, i.e., $\omega \geq \omega_T$ and
$\Lambda \leq \Lambda_T$; we choose (and other similar choices give equally
consistent results) thresholds $\omega_T = 0.5\omega'$ and $\Lambda_T
= 0.1\Lambda'$.  This criterion allows us to define a \textit{mask} which, when applied 
to the vorticity field of active flows, generates a filtered field
$\widetilde \omega = \omega$ in streaks and 0 otherwise.
In Fig.~\ref{ConditionedPdf}(c) we show $ \widetilde \omega$ for the same
flow realisation as in panel (a) to demonstrate the accuracy of this criterion
which goes beyond the binary classification of the Okubo-Weiss parameter and
picks out the streaky regions.

With the definitions of the flow topologies in place, we
can now unambiguously separate trajectories based on the flow regions they
encounter. An obvious quantification is the joint distribution of tracer
displacements $s$ with residence times $t$ in and out of streak regions.
Since the streaky regions occupy a relatively small
area-fraction $A_f$ of the flow, it is useful to look at the effective displacements $\phi \equiv
s/\sqrt{A_f}$  within a streak, and conversely $\phi \equiv
s/(1-\sqrt{A_f})$ outside it, to ascertain the
relative degree to which streaky and non-streaky regions assist persistent motion.

While streaks form a small fraction ($\sim 12\%$) of the flow resulting in a
relatively smaller residence time, tracers inside the
(Fig.~\ref{ConditionedPdf}(d)) streaks are advected much farther within this
short time than those outside (Fig.~\ref{ConditionedPdf}(e)). The joint
$\phi - t$ distribution is much wider when outside these
special structures, indicating that large residence times (in purely vortical
and straining patches) do not lead to large displacements. Thus, coherent,
persistent motion is intrinsically correlated with the special flow patterns
(Fig.~\ref{ConditionedPdf}(c)) that a bacterial suspension can spontaneously
generate.

\section*{First Passage Problem and Dynamical Heterogeneity}
This surprising ability to exploit self-generated patterns and their emergent hydrodynamics to aid persistent motion 
accords individuals a distinct advantage to reach targets---for nutrition, for
example---on much shorter time scales than it would be otherwise possible.
While this was certainly implicit in earlier observations of anomalous
diffusion through measurements of the mean-square-displacements~\cite{Sid21}, a direct
way of looking at this is to ask how such activity-induced streaks enhance the
efficacy of tracers in reaching distances $R$ away. This, of course, is the much
studied question in statistical physics of First Passage Problems~\citep{Chandra43,Majumdar13,Redner01,Bala21,Metzler14}, also used to quantify biological motility~\citep{chou2014first,bisht2017twitching,teimouri2019theoretical}: What is the
statistics of the time $t_R$ for tracers to reach (for the first time and be ``absorbed'' at) the
boundary of a circle of radius $R$ as illustrated in Fig.~\ref{Traj}; see also \url{https://youtu.be/CPJ3ZlXBf-k}.

In the context of bacterial suspensions and living systems in general, the
first passage characteristics are critical to how efficiently bacteria forage
for food and survive. In an active turbulent flow, whose evolution is 
governed by Eq.~\eqref{GNS}, a bacterium foraging for food can be thought 
of as a point tracer being advected until it 
reaches the food source at a distance $R$, at which point the search is concluded.  

\textcolor{black}{The probability of a tracer reaching (and being absorbed by) a boundary $R$ at
time $t = t_R$ when it starts from $r = 0$ at $t=0$ is given by the first
passage probability distribution $P(R,t_R)$ which is calculated from the
(survival) probability $S(t) = \int_{0}^{R} p(r,t) 2\pi r dr$ of not reaching
the boundary in time $t_R$ via $P(R,t_R) = -\frac{\partial S}{\partial t} \big
|_{t_R}$~\citep{Bala21,Seigert51}. The survival probability assumes isotropy of the distribution
$p(r,t)$ of tracers which, for low activities, satisfies the diffusion
equation~\citep{Risken96} with absorbing boundary condition $p(R,t) = 0$ and
the initial condition $p(r,0) = \delta (r)/2\pi r$. We begin by writing the Fokker-Planck equation in two-dimensions:
\begin{align}
	\partial_t p(r,t) = K_d\nabla^2 p(r,t)
	\label{Diff}
\end{align}
$K_d$ being the diffusion constant. Assuming a solution of the form $p(r,t) = u(r)f(t)$ yields separate differential equations for the temporal and radial parts to obtain:
\begin{align}
	u(r) &= J_0(Cr) = \sum^{\infty}_{n=0}\frac{(-1)^n}{(n!)^2}\left(\frac{Cr}{2}\right)^{2n}  \\
	f(t) &= f_0e^{-C^2 K_d t} 
\end{align}
where $f_0 = f(t)|_{t=0}$, and $J_0$ is the Bessel function of the first kind and order zero. 
By using the absorbing boundary condition to solve for the radial part $u(r)$, the delta-function initial condition
to determine $f_0$ and the constraint that $S(t=0) = 1$, we obtain 
\begin{equation}
	p(r,t) = \frac{1}{2\pi R^2 \left( \sum_i \frac{J_1(h_i)}{h_i}\right)} \sum_i J_0\left(\frac{h_i r}{R}\right) e^{-h_i K_dt/R^2}; 
\end{equation}
thence the survival probability
\begin{equation}
S(t) = \int_{0}^{R}p(r,t) 2 \pi r dr = \sum_i e^{-h_i K_dt/R^2} \frac{J_1(h_i)}{h_i}.
\end{equation}
Here, $J_1$ is the  Bessel function of the first kind and order one, whereas $h_i$ are the zeros of $J_0(CR)$. 
The first passage time distribution is then simply 
\begin{equation}
	P(R,t_R) = -\partial_t S|_{t=t	_R} = \frac{K_d}{R^2} \sum_i e^{-h_i K_dt_R/R^2} J_1(h_i)
\end{equation}
The dominant contribution to the first passage distribution comes from the smallest $h_i=A$ and
yields the well-known result: $P(R,t_R) \sim  \frac{K_d}{R^2} e^{-A K_d t_R/R^2}$
~\citep{Akshay2020}.} The first passage distribution can be generalised to the case of
anomalous diffusion as $P(R,t_R) \sim e^{-t_R/R^{2/\xi}}$, where the scaling
exponent $1 \le \xi \le 2$ is a consequence of the anomalous diffusion in the
mean-square-displacement $\Delta x^2 \sim t^\xi$ (and which can also be
obtained analytically by solving the fractional Fokker-Planck
equation~\citep{Seigert51,RangaA2000,RangaE2000,Gitterman2000,Palyulin19}). 

\begin{figure*}
\includegraphics[width=1\textwidth]{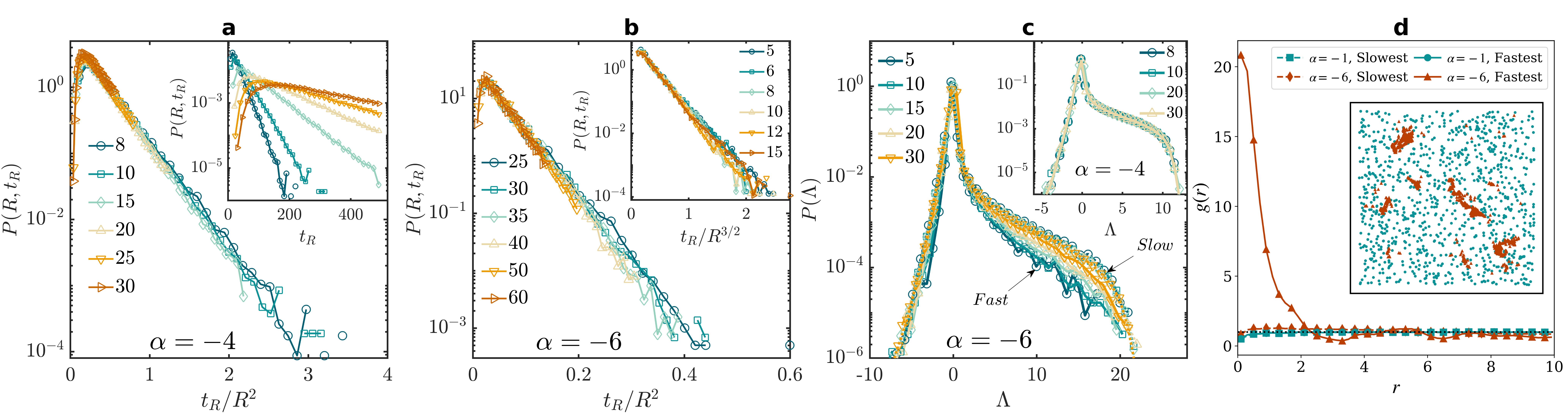}
	\caption{(a) Semi-log plots of the first passage distributions for
	mildly active suspensions collapse on a single curve for rescaled time $t_R/R^2$
	corresponding the diffusive transport for different values of $R$; the
	inset shows the same without the rescaling of time. (b) Analogous plots
	for highly active suspensions show a similar diffusion-driven collapse
	\textit{only} for large distances and time; for short distance (inset)
	the curves collapse when time is rescaled as  $t_R/R^{3/2}$ which
	factors in anomalous diffusion. (c) Probability distribution functions
	of $\Lambda$ conditioned on the fastest (solid-lines) and
	slowest (dashed-lines) $10\%$ of tracers for $\alpha = -6$ and $\alpha =
	-4$ (inset) and for various target radii $R$. The distributions differ
	only for the highly active suspension and for small $R$ showing the
	emergence of a true dynamical heterogeneity. This is visually
	illustrated in the inset of panel (d) showing that the origins of the
	fastest $1\%$ tracers for $\alpha=-1$ are randomly distributed whereas those
	for $\alpha=-6$ are strongly localised suggesting ``lucky'' spots for
	efficient motility. This is confirmed in (d) measurements of the pair-distribution
	function which gives a pronounced peak for the fastest tracer origins
	for $\alpha=-6$, while rapidly converging to the uniform distribution
	$g(r)=1$. Similar measurements for the fastest tracers for lower activity and indeed 
	for the slowest trajectories for all activity show uniform distribution throughout.}
\label{DynHet}
\end{figure*}

In the inset of Fig.~\ref{DynHet}(a) we show the first passage time
distributions for different values of $R$ which collapse on to a unique curve
when the first passage times are scaled as $t_R/R^2$
(Fig.~\ref{DynHet}(a)) consistent with our theoretical prediction for moderately active suspensions. On
making suspensions more active, as shown in Fig.~\ref{DynHet}(b), the diffusive 
$t_R/R^2$ scaling still remains approximately true but only for large values of
$R$. Indeed, for smaller values of $R$ (Fig.~\ref{DynHet}(b), inset), the
distributions collapse only when the first passage times are scaled as
$t_R/R^{3/2}$, accounting for the enhanced motility. This is because at such short distances, persistent trajectories
contribute to the statistics of first passage overwhelmingly and hence the
scaling exponent associated with anomalous diffusion in such systems, $\xi
\approx 4/3$~\cite{Sid21}, leads to an anomalous form of the first passage
distribution. 

While the normalized first passage distributions give a statistical sense for 
an \textit{ensemble} of trajectories, it does not allow us to have a sense of the variations, within an ensemble, 
of individual trajectories. Since tracers sample
the entire phase space, the issue of an incipient \textit{dynamical heterogeneity} in the flow is a vexing one.

In order to understand this, we take a fraction ($10\%$, for sufficient statistics) of the fastest and slowest trajectories that reach
various target radii $R$, for different values of $\alpha$. Visually, the
fastest and slowest trajectories are different: The former mostly
straight and persistent, while the latter convoluted and meandering for \textit{all}
levels of activity. However, the Lagrangian history of these tracers, i.e., the
$\Lambda$ values they encounter before hitting their targets, shows, as illustrated 
in the inset of Fig.~\ref{DynHet}(c) that at
mild levels of activity there is no distinction between the
underlying Eulerian field sampled by the fastest and slowest tracers.
These conditional distributions of the Okubo-Weiss parameter show that the
heterogeneity in trajectories, up to intermediate levels of activity, is simply
a statistical consequence of the \textit{random} sampling of the flow. 
However, for highly active suspension, where anomalous diffusion becomes robust,
signatures of dynamical heterogeneity, as seen in the clearly different Lagrangian histories of the
fastest and slowest trajectories (see Fig.~\ref{DynHet}(c), manifest themselves. Consistent with the
findings so far that persistent motion favours quiescent field regions, the
fastest tracers sample milder regions of the Okubo-Weiss field, in comparison
to the slowest tracers. This distinction, moreover, is starker for smaller $R$
values, while at large values of $R$, the tracers begin to experience (statistically) similar
underlying fields. 

This evidence for dynamical heterogeneity in the flow, such that the variation
in the nature of trajectories is not \textit{merely} statistical as happens in inertial turbulence~\citep{scatamacchia2012extreme}, is further
strengthened by looking at where in the flow do the fastest and slowest tracers
originate, for a given value of $R$. In other words, do trajectories get
\textit{lucky} by being at the right place at the right time, which allows
them to reach their targets faster than others?  This would suggest that for
highly active suspensions, the initial locations of these fast, lucky
trajectories must be clustered in nearby regions while for less active
flows they are more uniformly distributed. In the inset of
Fig.~\ref{DynHet}(d), we show a plot of the starting points of the fastest $1\%$ of the trajectories for
a target at $R=15$. We indeed find that for the highly active suspensions ($\alpha
= -6$) these points remain strongly clustered, unlike the random and uniform
distribution seen for the less active ($\alpha = -1$) case. While confirming a
dynamically heterogeneous flow, this also explains the bundling of these trajectories already seen
in Fig.~\ref{Traj}(b), where the fastest tracers (after their initial locations
are artificially superimposed at the center of the circle), evolve as coherent
\textit{bundles}, arriving at their target circle and forming a very
anisotropic distribution around the circumference. This is because the bundles
originate clustered together in a few special locations in the flow and hence
follow very similar trajectories.  We quantify the degree of this clustering by
calculating the pair-distribution function $g(r)$, for the fastest and slowest $1\%$ of the tracers, in Fig.~\ref{DynHet}(d). The $g(r)$ of the origins of the slowest
tracers quickly attains a value close to $1$, irrespective of activity,
corresponding to a uniform (or ideal-gas) distribution of points (albeit with a
small overshoot for $\alpha=-6$ at small $r$, showing weak preferential
clustering also for the slowest tracers). The fastest tracer origins for high
activity show a pronounced peak in the pair distribution at small values of
$r$, reflecting the presence of these lucky spots.  At mild activity, the
$g(r)$ for the fastest tracer origins rapidly attains to a value of $1$ as
well, reflecting the fact that the heterogeneity of trajectories is merely statistical.  

\section*{Pair-Dispersion}
So far we have focussed our attention on single-particle statistics. But before we 
conclude, we must consider the implications of these anomalies on the Richardson-Obukhov
pair-dispersion problem~\citep{richardson1926,obukhov1941distribution} which
informs much of our understanding of inertial
turbulence~\citep{salazar2009two}.
Pair-dispersion is simply the statistics of separation of particle pairs that originate within a small distance $\epsilon$ of each other measured 
as $\left \langle r_p^2(t) \right\rangle = \left \langle | {\mathbf{x}_1(t)} - {\mathbf{x}_2(t)} | ^2\right\rangle$, 
where $\mathbf{x}_1(t)$ and $\mathbf{x}_2(t)$ are the positions of a particle pair, with $|\mathbf{x}_1(0)-\mathbf{x}_2(0)| = \epsilon$ being 
their initial separation, and $\ang{\cdot}$ denotes averaging over all particle pairs. 

\begin{figure*}
\includegraphics[width=1\textwidth]{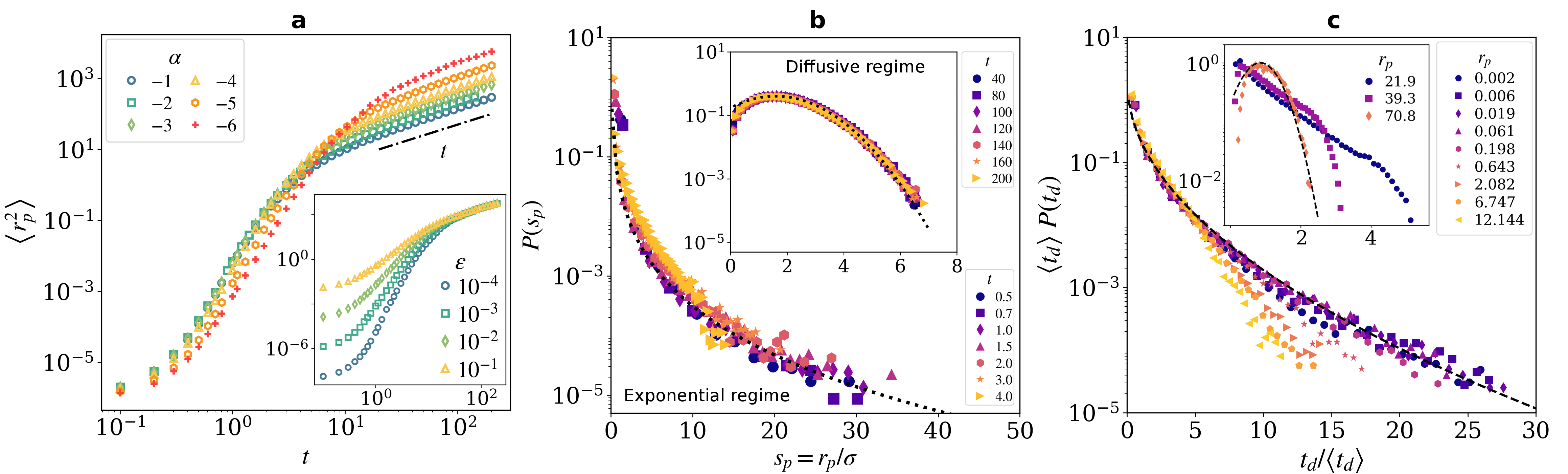}
	\caption{(a) The dependence of the relative pair separation on $\alpha$ (with $\epsilon=0.001$) and (inset) on $\epsilon$ (for $\alpha = -6$). 
	(b) Probability distribution of the normalized pair separation $s_p$ follows a stretched exponential for times up to diffusive pair-separation, 
	whence (inset) the distribution becomes Gaussian. (c) The probability distribution of pair-separation doubling times $t_d$ 
	are also stretched exponentials at small separations; the curves become steeper with increasing $r_p$ and (inset) converge to a Gaussian distribution for large $r_p$.}
\label{RSon}
\end{figure*}

Figure~\ref{RSon}(a) shows pair-separation for increasingly active suspensions,
with $\epsilon=0.001$. Interestingly, the influence of activity appears to be
mild, with only a slight change in the extent of the intermediate scaling; even
for $\alpha=-6$, a scaling regime seems to extend only up to a decade. The limited extent of scaling, and
its significant deviation from the Richardson prediction $\sim t^3$ for inertial turbulence, may stem from the lack of an
inertial range in active turbulence. It has been shown that active turbulence,
at least using the nematohydrodynamic model and other minimal models for wet
active nematics~\citep{urzay2017multi,carenza2020cascade,alert2021active}, does
not have an energy cascade resulting from non-linear mode interactions. While,
admittedly, our model differs from these, it is likely that a similar situation
arises here. In fact, the presence of a particularly wide inertial range and
scale separation (effectively, a large Reynolds number) is essential to
observe the Richardson scaling regime even in inertial turbulence, despite
cascade dynamics, and persistent motion in trajectories (ballistic L\'evy
walks) can further lead to deviation from Richardson
scaling~\citep{Boffetta02}. While there is no natural way to increase scale
separation in our system, we tested various domain sizes for $\alpha=-6$, from
$L=20$ (on $N=1024$) to $L=80$ (on $N=4096$), and obtained identical
pair-separation curves, without an increase in the extent of scaling. The inset
of Fig.\ref{RSon}(a) shows the influence of the initial separation
$\epsilon$ on pair-separation for $\alpha=-6$. As $\epsilon$ decreases, the
slope of the intermediate range increases significantly. This $\epsilon$
dependence is consistent with findings from two-dimensional inertial
turbulence~\citep{rivera2005pair,xia2019local}, and the apparent steepening of
the slope is a consequence of an exponential initial separation tending to
diffusive separation at longer times.

In the absence of robust Richardson scaling, other pair-separation measures
have been proposed to quantify dispersion
statistics~\citep{jullien1999richardson,Boffetta02}.  One such is the
probability distribution of pair-separations $r_p$, at different times during
the growth of $\left\langle r_p^2 \right\rangle$. Figure~\ref{RSon}(b) shows
the probability distribution of the rescaled (with the variance)
pair-separations $s_p = r_p/\sigma$, at various time instances, for
trajectories with an initial separation of $\epsilon = 0.001$ and for
$\alpha=-6$. The separations $s_p$ collapse to a single stretched-exponential
distribution while at very long times, well into the diffusive pair-separation
regime $\left\langle r_p^2\right\rangle \sim t$, the $s_p$ distribution becomes
Gaussian. All this is consistent with findings from two-dimensional inertial
turbulence~\citep{jullien1999richardson,rivera2005pair}, where a
stretched-exponential and Gaussian distribution of rescaled separations is
found for the forward enstrophy-cascade and inverse energy-cascade regimes,
respectively. This shows that the pair-separation process is self-similar in
time, following different distributions during the rapid and (eventually)
diffusive growth. 

Related to the first passage problem, we consider the distribution of separation
doubling-times~\citep{Boffetta02,rivera2005pair}. This is defined as the time
$t_d$ it takes for an initial separation $r_p$ to grow to a scale $\rho r_p$.
Here, we can also expect an influence of the persistence in trajectories.
Smaller initial separations $r_p$ will lead to longer correlated motion, as
trajectory-pairs will sample similar flow geometries (Fig.~\ref{RegionalTraj}).
This would lead to relatively longer
doubling times, and a wider $t_d$ distribution. At large initial separations,
where trajectory-pairs are essentially uncorrelated to begin with, the
separation doubling will occur more rapidly. This is precisely what is
observed in Fig.~\ref{RSon}(c) for $\rho=1.8$; our 
results are similar for other $\rho$ values that are not too small. The
distributions of doubling-times $t_d$, normalized by the mean doubling time
$\ang{t_d}$, for small values of $r_p$, follow a stretched exponential function, similar
to two-dimensional inertial turbulence in the enstrophy-cascade
range~\citep{rivera2005pair}. The distributions become steeper with increasing
$r_p$, reflecting the decorrelation within trajectory
pairs. The inset of Fig.~\ref{RSon}(c) shows that at very large values of $r_p$,
deep into the diffusive pair-separation phase, the distribution of $t_d$, as anticipated, approaches a 
Gaussian.

\section*{Conclusions}
\corr{Anomalous diffusion and L\'evy walks are common place in a wide variety of biological systems. However,
in dense suspensions of microorganisms, such as bacterial swarms, which are a typical example
of flowing active matter and a testing ground for active turbulence, the experimental evidence
for such phenomena is only recent~\citep{ariel2015swarming}. Furthermore,
theoretical studies have been unable to go beyond the simple diffusion picture and predict, detect
or understand the emergence of such super-diffusive regimes in active turbulence. This issue was resolved by a recent work~\citep{Sid21} showing that anomalous Lagrangian statistics manifest only in extremely active suspensions, with $\alpha \sim -6$. Further exploring such highly active suspensions consistently revealed signatures of anomalies in Lagrangian measurements. The streaky flow regions, together with an emergent dynamical heterogeneity, contrive to selectively propel tracers persistently and aid in anomalous first-passage statistics. The results presented in this work should be readily amenable to experimental measurements in super-diffusive bacterial swarms, as well as in a diverse class of active flows. Whether the conjoined emergence of Eulerian streaks at high activity, and their role in promoting persistent motion, is universal, remains to be ascertained, and the true mechanisms for manifesting Lagrangian anomalies may well be system dependent. Notwithstanding, we feel that our results show a biologically crucial aspect of the generalized hydrodynamics model beyond what has been previously observed, extending its applicability to a realm closer to experimental observations.}

\corr{Further, }measures like pair-dispersion mirror fundamental quantities like
Lyapunov exponents, that characterize chaos in both diverging trajectories and
solutions of the hydrodynamic equations. For instance, the flow Lyapunov
exponent is known to increase with Reynolds number in inertial turbulence
approximately as $\sim
\mathrm{Re}^{1/2}$~\citep{mukherjee2016predictability,boffetta2017chaos,berera2018chaotic}.
The analogous effect of activity on Lyapunov exponents in active turbulence
remains unknown and is part of ongoing work, both within the
Eulerian~\citep{mukherjee2016predictability,boffetta2017chaos,berera2018chaotic}
and Lagrangian framework~\cite{Bec2006,Ray2018}, which looks at different aspects of
many-body chaos~\citep{murugan2021many} in such systems. Indeed, a complete
understanding of the dynamical facets of active turbulence calls for a complete
characterization of the skeleton of chaos underlying active flows.
\corr{It is interesting to note that a recent work~\citep{Pandit22} on the (Lagrangian) irreversibility 
in such suspensions shows similar anomalies between active and inertial turbulence.}
Lastly, our work also lays out a systematic framework for the Lagrangian
analysis of active flows, including measures for identifying geometrical
structures emerging in the Eulerian fields. \corr{Adopting these in future work} will not only ascertain the extent of universality of
active turbulence anomalies, but also help connect hydrodynamic theories of
active matter more closely to the physical picture. 

\begin{acknowledgements} We thank Martin James for several useful discussions. The simulations were performed on the ICTS clusters
{\it Tetris}, and {\it Contra}. SSR acknowledges
	SERB-DST (India) project DST (India) project MTR/2019/001553 and STR/2021/000023 for
financial support.  The authors acknowledges the support of the DAE, Govt. of
India, under project no.  12-R\&D-TFR-5.10-1100 and project no.  RTI4001
\end{acknowledgements}

\bibliography{ref_act_turb}

\end{document}